\documentclass[journal,onecolumn,12pt]{IEEEtran}
\renewcommand{\baselinestretch}{1.5}
\usepackage{cite}
\usepackage{amssymb}
\usepackage{amsmath}
\usepackage{graphicx}
\usepackage{epsfig}
\usepackage{amsthm}
\usepackage{subfigure}
\usepackage{algorithmic}

\begin{document}
\title{Optimal Coding for the Erasure Channel with Arbitrary Alphabet Size}
\author{\authorblockN{Shervan Fashandi, Shahab Oveis Gharan and Amir K. Khandani\\}\authorblockA{ECE Dept., University of Waterloo, Waterloo, ON, Canada, N2L3G1\\ email: \{sfashand,shahab,khandani\}@cst.uwaterloo.ca}}
\vskip 0cm
\maketitle
\vskip 0cm
\begin{abstract}
An erasure channel with a fixed alphabet size $q$, where $q\gg 1$,
is studied . It is proved that over any erasure channel (with or
without memory), \textit{Maximum Distance Separable} (MDS) codes
achieve the minimum probability of error (assuming maximum
likelihood decoding). Assuming a memoryless erasure channel, the
error exponent of MDS codes are compared with that of random codes 
and linear random codes. It is shown that the envelopes of all these 
exponents are identical for rates above the critical rate. Noting the 
optimality of MDS codes, it is concluded that both random codes and linear 
random codes are exponentially optimal, whether the block sizes is larger or smaller 
than the alphabet size. \footnote{Financial support provided by Nortel 
and the corresponding matching funds by the Natural Sciences and Engineering 
Research Council of Canada (NSERC), and Ontario Centres of Excellence (OCE) are 
gratefully acknowledged.}
\end{abstract}

\section{Introduction}
\label{section:Introduction} Erasure channels with large alphabet
sizes have recently received significant attention in networking
applications. Different erasure channel models are adopted to study
the performance of end-to-end connections over the
Internet~\cite{Zakhor2001,Dairaine2005}. In such models, each packet
is seen as a $q=2^b$-ary symbol where $b$ is the packet length in
bits. In this work, a memoryless erasure channel with a fixed, but
large alphabet size is considered. The error probability over this
channel (assuming maximum-likelihood decoding) for \textit{Maximum Distance 
Separable} (MDS) and random codebooks are compared and shown to be 
exponentially identical for rates above the critical rate.

Shannon~\cite{Shannon1948} was the first who observed that the error
probability for maximum likelihood decoding of a random code
($P_{E,M\!L}^{r\:\!\!a\:\!\!n\:\!\!d}$) can be upper-bounded by an
exponentially decaying function with respect to the code block
length $N$. This exponent is positive as long as the rate stays
below the channel capacity, $R < C$ . Following this 
result, tighter bounds were proposed in the literature \cite{Elias1955,
Shannon1967,Gallager1968ErrExp}. For rates below the critical rate, modifications 
of random coding are proposed to achieve tighter 
bounds~\cite{Forney1968}. Interestingly, the exponential upper-bound on
$P_{E,M\!L}^{r\:\!\!a\:\!\!n\:\!\!d}$ remains valid regardless of
the alphabet size $q$, even in the case where $q$ is larger than the
block size $N$ (e.g. see the steps of the proofs in
\cite{Gallager1968ErrExp}). There is also a lower-bound on the probability
of error using random coding which is known as the sphere packing
bound~\cite{Gallager1968SPB}. For channels with a relatively small
alphabet size ($q \ll N$), both the sphere packing lower-bound and the
random coding upper-bound on the error probability are exponentially
tight for rates above the critical rate~\cite{Gallager1973}. However, the sphere packing
bound is not tight if the alphabet size, $q$, is comparable to the
coding block length $N$ (noting the terms $o_1(N)$ and $o_2(N)$ in~\cite{Gallager1968SPB}). 

Probability of error, minimum distance, and distance distribution of random linear codes are 
discussed in~\cite{Pierce1967,Barg2002}. Pierce studies the asymptotic behavior of 
the minimum distance of binary random linear codes~\cite{Pierce1967}. Error exponent of random linear 
codes over a binary symmetric channel is analyzed in~\cite{Barg2002}. Barg et al. also study the minimum 
distance and distance distribution of random linear codes and show that random linear codes have better 
expurgated error exponent as compared to random codes for rates below the critical rate~\cite{Barg2002}.

\textit{Maximum Distance Separable} (MDS)~\cite{Roth2006MDS} codes are
optimum in the sense that they achieve the largest possible minimum
distance, $d_{min}$, among all block codes of the same size. Indeed,
any codeword in an MDS code of size $[N, K]$ can be successfully decoded from any subset of
its coded symbols of size $K$ or more. This property makes MDS codes
suitable for use over erasure channels like the
Internet~\cite{Zakhor2001,Dairaine2005,Peng2005}. However, the
practical encoding-decoding algorithms for such codes have quadratic
time complexity in terms of the code block length~\cite{Alon1995}.
Theoretically, more efficient ($O\left( N\log^2 N \right)$) MDS
codes can be constructed based on evaluating and interpolating
polynomials over specially chosen finite fields using Discrete
Fourier Transform~\cite{Justesen1976}. However, in practice these
methods can not compete with the quadratic methods except for
extremely large block sizes. Recently, a family of almost-MDS codes
with low encoding-decoding complexity (linear in length) is proposed
and shown to provide a practical alternative for coding over the
erasure channels like the Internet~\cite{Luby2001}. In these codes,
any subset of symbols of size $K(1+\epsilon)$ is sufficient to
recover the original $K$ symbols with high probability
\cite{Luby2001}. Fountain codes, based on the idea of
almost-MDS codes with linear decoding complexity, are proposed for information multicasting to many
users over an erasure channel \cite{Luby2002, Shokrollahi2006}.

In this work, a memoryless erasure channel with a fixed, but large
alphabet size is studied. First, it is proved that MDS block codes
offer the minimum probability of decoding error over any erasure
channel. Then, error exponents of MDS codes, random codes, and linear 
random codes for a memoryless erasure channel are analyzed and shown 
to be identical for rates above the critical rate. Combining the two 
results, we conclude that both random codes and linear random codes are 
exponentially as good as MDS codes (exponentially optimal) over a wide 
range of rates.

The rest of this paper is organized as follows. In
section~\ref{section:ErasureChModel}, the erasure channel model is
introduced, and the assumption of large alphabet sizes is justified.
Section \ref{section:OptMDS} proves that MDS codes are optimum over
any erasure channel. Error exponents of MDS codes, random codes, and linear 
random codes over a memoryless erasure channel are compared in
section~\ref{section:ErrExpMDSrandom}. Finally, section~\ref{section:Conclusion} concludes the paper.

\begin{figure}
    \centering
    \includegraphics[scale=0.35]{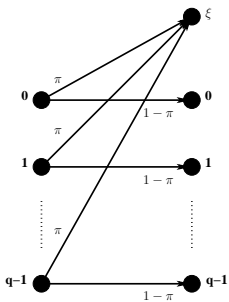}
    \caption{Erasure memoryless channel model with the alphabet size $q$, probability of erasure $\pi$, and the erasure symbol $\xi$.}
    \label{fig:ErasureMemorylessCh}
\end{figure}

\section{Erasure Channel Model}
\label{section:ErasureChModel} The  memoryless erasure channel
studied in this work has the alphabet size $q$ and the erasure
probability $\pi$ (see Fig.~\ref{fig:ErasureMemorylessCh}). The
alphabet size $q$ is assumed to be fixed and large, i.e., $q \gg 1$.

The described channel model occurs in many practical scenarios such
as the Internet. From an end to end protocol's perspective,
performance of the lower layers in the protocol stack can be modeled
as a random \textit{channel} called an \textit{Internet channel}.
Since each packet usually includes an internal error detection
mechanism (for instance a Cyclic Redundancy Check), the Internet
channel can be modeled as an erasure channel with packets as
symbols~\cite{Fashandi2007Glob}. If each packet contains $b$ bits,
the corresponding channel will have an alphabet size of $q=2^b$
which is huge for typical packet sizes. Therefore, in practical
networking applications, the block size is usually much smaller than
the alphabet size. Algebraic computations over Galois fields
$\mathbb{F}_q$ of such large cardinalities is now practically
feasible with the increasing processing power of electronic
circuits. Note that network coding schemes, recently proposed and
applied for content distribution over large networks, have a
comparable computational complexity~\cite{Koetter2003, Ho2003, Chou2003, Gkantsidis2005,Gkantsidis2006, Ho2006, Jaggi2007}.

Note that all the known MDS codes have alphabets of a large size
(growing at least linearly with the block length $N$). Indeed, a
conjecture on MDS codes states that for every linear $[N, K]$ MDS
code over the Galois field $\mathbb{F}_q$, if $1 <K < q$, then
$N\leq q+1$, except when $q$ is even and $K=3$ or $K=q-1$, for which
$N \leq q+2$~\cite{Walker1996}. To have a feasible MDS code over a
channel with the alphabet size $q$, the block size $N$ should
satisfy $N \leq q+1$. 


\section{Optimality of MDS Codes over Erasure Channels}
\label{section:OptMDS}

\textit{Maximum Distance Separable} (MDS) codes are optimum in the
sense of achieving the largest possible minimum distance, $d_{min}$,
among all block codes of the same size~\cite{Roth2006MDS}. The
following proposition shows that MDS codes are also optimum over any
erasure channel in the sense of achieving the minimum probability of
decoding error.

\textbf{Definition I.} An erasure channel is defined as the one which
maps every input symbol to either itself or to an erasure symbol
$\xi$. More accurately, an arbitrary channel (memoryless or with memory) with the input 
vector $\mathbf{x}\in\mathcal{X}^N$, $|\mathcal{X}|=q$ , the output 
vector $\mathbf{y}\in \left(\mathcal{X} \cup \{\xi\}\right)^N$, and the transition
probability $p\left(\mathbf{y}|\mathbf{x}\right)$ is defined to be erasure \textit{iff} it satisfies the 
following conditions:
\begin{enumerate}
\item $p\left( y_j \notin \left\{ x_j,\xi \right\} \right | x_j)=0,~\forall~j$, where $x_j$, $y_j$, and $e_j$ denote the $j$'th elements of the vectors $\mathbf x$, $\mathbf y$, and $\mathbf e$.
\item Defining the erasure identifier vector $\mathbf e$ as
\begin{equation}
e_j = \left \{ \begin{array}{ll}
1 & y_j = \xi \\
0 & \mbox{otherwise} \end{array} \right. \nonumber
\end{equation}
$p(\mathbf{e} | \mathbf{x})$ is independent of $\mathbf x$.
\end{enumerate}

\textbf{Proposition I.}  A block code of size $[N,K]$ with equiprobable codewords over an arbitrary erasure 
channel (memoryless or with memory) has the minimum probability of error (assuming optimum, i.e.,
maximum likelihood decoding) among all block codes of the same size
\textit{if} that code is \textit{Maximum Distance Separable} (MDS).

\textbf{Proof.} Consider a $[N,K,d]$ codebook $\mathcal{C}$ with the
$q$-ary codewords of length $N$, number of code-words $q^K$, and
minimum distance $d$. The distance between two codewords is defined
as the number of positions in which the corresponding symbols are
different (Hamming distance). A codeword $\mathbf{x} \in
\mathcal{C}$ is transmitted and a vector $\mathbf{y} \in
\left(\mathcal{X} \cup \{\xi\}\right)^{N}$ is received. The number
of erased symbols is equal to the Hamming weight of $\mathbf{e}$
denoted by $w(\mathbf{e})$. An error occurs if the decoder decides
for a codeword different from $\mathbf{x}$. Let us assume that the
probability of having a specific erasure pattern $\mathbf{e}$ is
$\mathbb{P}\{\mathbf{e}\}$ which is independent of the transmitted
codeword (depends only on the channel). We assume a specific erasure
vector $\mathbf{e}$ of weight $m$. The decoder decodes the
transmitted codeword based on the $N-m$ correctly received symbols.
We partition the code-book, $\mathcal{C}$, into $q^{N-m}$ bins, each
bin representing a specific received vector satisfying the erasure
pattern $\mathbf{e}$. The number of codewords in the $i$'th bin is
denoted by $b_{\mathbf{e}}(i)$ for $i=1,...,q^{N-m}$. Knowing the
erasure vector $\mathbf{e}$ and the received vector $\mathbf{y}$,
the decoder selects the bin $i$ corresponding to $\mathbf{y}$. The
set of possible transmitted codewords is equal to the set of
codewords in bin $i$ (all the codewords in bin $i$ are equiprobable
to be transmitted). If $b_{\mathbf{e}}(i)=1$, the transmitted
codeword $\mathbf{x}$ can be decoded with no ambiguity. Otherwise,
the optimum decoder randomly selects one of the
$b_{\mathbf{e}}(i)>1$ codewords in the bin. Thus, the probability of
error is $1-\frac{1}{b_{\mathbf{e}}(i)}$ when bin $i$ is selected.
Bin $i$ is selected if one of the codewords it contains is
transmitted. Hence, probability of selecting bin $i$ is equal to
$\frac{b_{\mathbf{e}}(i)}{q^K}$. Based on the above arguments,
probability of decoding error for the maximum likelihood decoder of
any codebook,$\mathcal{C}$, is equal to
\begin{eqnarray}
P_{E,M\!L}^{\mathcal{C}} & \hspace{-0.3cm} \stackrel{(a)}{=} & \hspace{-0.4cm}
\sum_{m=d}^{N}~\sum_{\mathbf{e}:w(\mathbf{e})=m}\mathbb{P}\{\mathbf{e}\}\mathbb{P}\{\mbox{error}|\mathbf{e}\} \nonumber \\
&\hspace{-0.3cm} = & \hspace{-0.4cm} \sum_{m=d}^{N}~\sum_{\mathbf{e}:w(\mathbf{e})=m}\mathbb{P}\{\mathbf{e}\}\sum_{i=1,~b_{\mathbf{e}}(i)>0}^{q^{N-m}}\left(1-\dfrac{1}{b_{\mathbf{e}}(i)}\right)\dfrac{b_{\mathbf{e}}(i)}{q^K} \nonumber \\
& \hspace{-0.3cm} \stackrel{(b)}{=} & \hspace{-0.4cm} \sum_{m=d}^{N}~\sum_{\mathbf{e}:w(\mathbf{e})=m}\mathbb{P}\{\mathbf{e}\}~\left(1-\dfrac{b_{\mathbf{e}}^+}{q^K}\right) \nonumber \\
&\hspace{-0.3cm} \stackrel{(c)}{\geq} & \hspace{-0.4cm} \sum_{m=d}^{N}~\sum_{\mathbf{e}:w(\mathbf{e})=m}\mathbb{P}\{\mathbf{e}\}~\left(1-\dfrac{\min\{q^K,q^{N-m}\}}{q^K}\right)
\label{equation:PEupperboundMDS}
\end{eqnarray}
where $b_{\mathbf{e}}^+$ indicates the number of bins containing one
or more codewords. $(a)$ follows from the fact that the transmitted
codeword can be uniquely decoded if the number of erasures in the
channel is less than the minimum distance of the codebook, and $(b)$
follows from the fact that
$\sum_{i=1}^{q^{N-m}}{b_{\mathbf{e}}(i)}=q^K$. $(c)$ is true since 
$b_{\mathbf{e}}^+$ is less than both the total number of codewords and the number of bins.

According to~\eqref{equation:PEupperboundMDS},
$P_{E,M\!L}^{\mathcal{C}}$ is minimized for a code-book $\mathcal{C}$
\textit{if} two conditions are satisfied. First, the minimum
distance of $\mathcal{C}$ should achieve the maximum possible value,
i.e., $d=N-K+1$. Second, we should have $b_{\mathbf{e}}^+=q^{N-m}$
for all possible erasure vectors $\mathbf{e}$ with any weight $d
\leq m \leq N$. Any MDS code satisfies the first condition by
definition. Moreover, it is easy to show that for any MDS code, we
have $b_{\mathbf{e}}(i)=q^{K-N+m}$. We first prove this for the case
of $m=N-K$. Consider the bins of an MDS code for any arbitrary
erasure pattern $\mathbf{e}, w(\mathbf{e})=N-K$. From the fact that
$d=N-K+1$ and $\sum_{i=1}^{q^K}{b_{\mathbf{e}}(i)}=q^K$, it is
concluded that each bin contains exactly one codeword. Therefore,
there exists only one codeword which matches any $K$ correctly
received symbols. Now, consider any general erasure pattern
$\mathbf{e}, w(\mathbf{e})=m>N-K$. For the $i$'th bin, concatenating
any $K-N+m$ arbitrary symbols to the $N-m$ correctly received
symbols results in a distinct codeword of the MDS codebook. Having
$q^{K-N+m}$ possibilities to expand the received $N-m$ symbols to
$K$ symbols, we have $b_{\mathbf{e}}(i)=q^{K-N+m}$. This completes
the proof$~\blacksquare$

\textbf{Remark I.} Proposition I is valid for any $N$ and $1 \leq K < N$. However, it does not guarantee the existence of an $[N,K]$ MDS code for 
all such values of $N$ and $K$. In fact, as stated in section~\ref{section:ErasureChModel}, a conjecture on MDS codes states that for every linear $[N, K]$ MDS
code over the Galois field $\mathbb{F}_q$, we have $N\leq q+1$ in most cases. Moreover, based on the Singleton bound, the inequality in~\eqref{equation:PEupperboundMDS} can 
be written as
\begin{equation}
P_{E,M\!L}^{\mathcal{C}} \geq  \sum_{m=N-K+1}^{N}~\sum_{\mathbf{e}:w(\mathbf{e})=m}\mathbb{P}\{\mathbf{e}\}~\left(1-\dfrac{q^{N-m}}{q^K}\right).
\label{equation:PEMLlowerbound} 
\end{equation}
Interestingly, this lower-bound is valid for any codebook $\mathcal{C}$ of size $[N,K]$, whether an MDS code of that size exists or not.

\textbf{Corollary I.} For $N \leq q+1$, converse of Proposition I is also true if the following condition is satisfied
\begin{equation}
\forall \mathbf{e}\in\{0,\xi\}^N:~\mathbb{P}\{\mathbf{e}\}>0
\label{equation:ConverseCondition} 
\end{equation}

\textbf{Proof.} For $N\leq q+1$ and $1 \leq K < N$, we know that an MDS code of size $[N,K]$ does exist (an $[N,K]$ Reed-Solomon code can be constructed 
over $\mathbb{F}_q$, see~\cite{Roth2006RSdec}). Let us assume the converse of Proposition I is not true. Then, there should be a non-MDS codebook, $\mathcal{C}$, with 
the size $[N,K,d]$, $d < N-K+1$, which achieves the minimum probability of error ($P_{E,M\!L}^{\mathcal{C}} = P_{E,M\!L}^{M\!DS}$). For any erasure vector $\mathbf{e}'$ with 
the weight $w(\mathbf{e}')=N-K$, we can write
\begin{eqnarray}
\mathbb{P}\{\mathbf{e}'\}~\left(1-\dfrac{b_{\mathbf{e}'}^+}{q^K}\right) & \stackrel{(a)}{\leq} &
\sum_{\mathbf{e}:w(\mathbf{e})=N-K}\mathbb{P}\{\mathbf{e}\}~\left(1-\dfrac{b_{\mathbf{e}}^+}{q^K}\right) \nonumber \\
& \stackrel{(b)}{\leq} & \sum_{m=d}^{N-K}~\sum_{\mathbf{e}:w(\mathbf{e})=m}\mathbb{P}\{\mathbf{e}\}~\left(1-\dfrac{b_{\mathbf{e}}^+}{q^K}\right) \nonumber \\
& \stackrel{(c)}{\leq} & \sum_{m=d}^{N-K}~\sum_{\mathbf{e}:w(\mathbf{e})=m}\mathbb{P}\{\mathbf{e}\}~\left(1-\dfrac{b_{\mathbf{e}}^+}{q^K}\right) \nonumber \\
& & + \sum_{m=N-K+1}^{N}~\sum_{\mathbf{e}:w(\mathbf{e})=m}
\mathbb{P}\{\mathbf{e}\}~\left(1-\dfrac{b_{\mathbf{e}}^+}{q^K}-1+\dfrac{q^{N-m}}{q^K}\right) \nonumber \\
& \stackrel{(d)}{=} & P_{E,M\!L}^{\mathcal{C}} - P_{E,M\!L}^{M\!DS}~\stackrel{(e)}{=}~0
\label{equation:bePrimePlus}
\end{eqnarray}
where $(a)$, $(b)$, and $(c)$ follow from the fact that $b_{\mathbf{e}}^+ \leq \min\{q^{N-m},q^K\}$ if $w(\mathbf{e})=m$. $(d)$ and $(e)$ are 
based on~\eqref{equation:PEupperboundMDS} and the assumption that $P_{E,M\!L}^{\mathcal{C}} = P_{E,M\!L}^{M\!DS}$. Combining~\eqref{equation:ConverseCondition} and~\eqref{equation:bePrimePlus} results 
in $b_{\mathbf{e}'}^+ = q^K$. Thus, we have $b_{\mathbf{e}'}(i)=1$ for all $1 \leq i \leq q^{K}$ and any $\mathbf{e}'$ with the weight of $w(\mathbf{e}')=N-K$. 

On the other hand, we know that the minimum distance of $\mathcal{C}$ is $d$. Thus, there exist two codewords $\mathbf{c}_1$ and $\mathbf{c}_2$ in $\mathcal{C}$ with the distance 
of $d$ from each other. We define the vector $\mathbf{e}_{12}$ as follows
\begin{equation}
\mathbf{e}_{12}=\left\{ \begin{array}{ll}
                        0 & \mbox{ if $\mathbf{c}_1=\mathbf{c}_2$} \\
                        1 & \mbox{ otherwise.}
                       \end{array} \right.
\end{equation}
It is obvious that $w(\mathbf{e}_{12})=d \leq N-K$. Then, we construct the binary vector $\mathbf{e}^{\star}$ by replacing enough number of zeros in $\mathbf{e}_{12}$ with ones such 
that $w(\mathbf{e}^{\star})=N-K$. The positions of these replacements can be arbitrary. In the binning corresponding to the erasure 
vector $\mathbf{e}^{\star}$, both $\mathbf{c}_1$ and $\mathbf{c}_2$ would be in the same bin since they have more than $K$ symbols in common. However, we know 
that $b_{\mathbf{e}^{\star}}(i)=1$ for all $1 \leq i \leq q^{K}$ since $w(\mathbf{e}^{\star})=N-K$. This contradiction proves the corollary$~\blacksquare$

The memoryless erasure channel obviously satisfies the condition in~\eqref{equation:ConverseCondition}. Combining Proposition I and Corollary I results in Corollary II.

\textbf{Corollary II.} A block code of size $[N,K]$ with equiprobable codewords over a memoryless erasure 
channel has the minimum probability of error (assuming optimum, i.e.,maximum likelihood decoding) among all block codes of the same size
\textit{iff} that code is \textit{Maximum Distance Separable} (MDS).

\subsection{MDS codes with Suboptimal Decoding}
\label{subsection:MDSsuboptimal} In the proof of proposition~I, it
is assumed that the received codewords are decoded based on maximum
likelihood decoding which is optimum in this case. However, in many
practical cases, MDS codes are decoded by simpler 
decoders~\cite{Roth2006RSdec}. Such suboptimal decoders can perfectly 
reconstruct the codewords of a $[N,K]$ codebook if they receive $K$ or 
more symbols correctly. In case more than $N-K$ symbols are erased, a 
decoding error occurs. Let $P_{E,s\:\!\!u\:\!\!b}^{M\!DS}$ denote the probability of
this event. $P_{E,s\:\!\!u\:\!\!b}^{M\!DS}$ is obviously different
from the decoding error probability of the maximum likelihood
decoder denoted by $P_{E,M\!L}^{M\!DS}$. Theoretically, an optimum
maximum likelihood decoder of an MDS code may still decode the
original codeword correctly with a positive, but small probability,
if it receives less than $K$ symbols. More precisely, according to
the proof of Proposition I, such a decoder is able to correctly
decode an MDS code over $\mathbb{F}_q$ with the probability of
$\frac{1}{q^i}$ after receiving $K-i$ correct symbols. Of course,
for Galois fields with large cardinality, this probability is
usually negligible. The relationship between
$P_{E,s\:\!\!u\:\!\!b}^{M\!DS}$ and $P_{E,M\!L}^{M\!DS}$ can be
summarized as follows
\begin{eqnarray}
P_{E,M\!L}^{M\!DS} & \hspace{-0.3cm} = & \hspace{-0.3cm} P_{E,s\:\!\!u\:\!\!b}^{M\!DS} - \hspace{-0.1cm} \sum_{i=1}^{K} \dfrac{\mathbb{P\{\mbox{$K-i$ symbols received correctly}\}}}{q^i}    \nonumber \\
                          & \hspace{-0.3cm} \geq & \hspace{-0.3cm} P_{E,s\:\!\!u\:\!\!b}^{M\!DS} - \hspace{-0.1cm} \sum_{i=1}^{K} \dfrac { \mathbb{P\{\mbox{$K-i$ symbols received correctly}\}} }{q}\nonumber \\
                          & \hspace{-0.3cm} = & \hspace{-0.3cm} P_{E,s\:\!\!u\:\!\!b}^{M\!DS} \left( 1-\dfrac{1}{q} \right).
\end{eqnarray}
Hence, $P_{E,M\!L}^{M\!DS}$ is bounded as
\begin{equation}
P_{E,s\:\!\!u\:\!\!b}^{M\!DS} \left( 1-\dfrac{1}{q} \right)  \leq  P_{E,M\!L}^{M\!DS}  \leq  P_{E,s\:\!\!u\:\!\!b}^{M\!DS}.
\label{equation:PEPEboundedML}
\end{equation}


\section{Error Exponents of MDS, Random, and Linear Random Codes}
\label{section:ErrExpMDSrandom}

\subsection{Error Exponent of MDS Codes over a Memoryless Erasure Channel}
\label{subsection:ErrExpMDSmemorylessErasureCh} 
Consider a block code of size $[N,K]$ over the memoryless erasure channel of
Fig.~\ref{fig:ErasureMemorylessCh}. Let $\alpha=\frac{N-K}{N}$
denote the coding overhead. For a $q$-ary $[N,K]$ code, the rate per
symbol, $R$, is equal to
\begin{equation}
R=\dfrac{K}{N} \log q = (1-\alpha) \log q.
\label{equation:RvsAlpha}
\end{equation}
In a block code of length $N$, the number of lost symbols would be
$\sum_{i=1}^{N}e_i$ where $e_i$ is defined in Proposition I. Thus,
the probability of decoding error for the suboptimal decoder of
subsection~\ref{subsection:MDSsuboptimal} can be written as
\begin{equation}
P_{E,s\:\!\!u\:\!\!b}^{M\!DS}=\mathbb{P}
\left\{ \dfrac{1}{N} \sum_{i=1}^{N}e_i > \alpha \right\} = \sum_{i=0}^{K-1} P_i
\label{equation:Binomial}
\end{equation}
where $P_i$ denotes the probability that $i$ symbols are received
correctly. Since $e_i$'s are i.i.d random variables with Bernoulli
distribution, we have $P_i=\left( 1-\pi \right)^i \pi^{N-i}
\binom{N}{i}$. It is easy to see that
\begin{eqnarray}
\dfrac{P_i}{P_{i-1}}=\dfrac{(N-i+1)(1-\pi)}{i\pi} > 1 & \mbox{for } i=1,\cdots,K-1 
\label{equation:PiPi_1}
\end{eqnarray}
if $\alpha=\frac{N-K}{N}>\pi$. According to equation
\eqref{equation:RvsAlpha}, the condition $\alpha>\pi$ can be
rewritten as $R< \left( 1-\pi \right) \log q = C$ where $C$ is
the capacity of the memoryless erasure channel. Therefore, the
summation terms in equation~\eqref{equation:Binomial} are always
increasing, and the largest term is the last one. Now, we can bound
$P_{E,s\:\!\!u\:\!\!b}^{M\!DS}$ as $P_{K-1} \leq
P_{E,s\:\!\!u\:\!\!b}^{M\!DS} \leq K P_{K-1}$. The term
$\binom{N}{K-1}$ in $P_{K-1}$ can be bounded using the fact that for
any $N>K>0$, we have~\cite{Cover1991}
\begin{equation}
\dfrac{1}{N+1} e^{N H \left(\frac{K}{N} \right) } \leq \binom{N}{K} \leq e^{N H \left(\frac{K}{N} \right) }
\label{equation:BinomBound}
\end{equation}
where the entropy, $H \left(\frac{K}{N} \right)$, is computed in
nats. Thus, $P_{E,s\:\!\!u\:\!\!b}^{M\!DS}$ is bounded as
\begin{equation}
\frac { \pi (1-\alpha) N e^{-Nu(\alpha)} } { (1-\pi) (N+1) (\alpha N+1) }
\hspace{-0.15cm} \leq P_{E,s\:\!\!u\:\!\!b}^{M\!DS} \leq \hspace{-0.15cm}
\frac { \pi (1-\alpha)^2 N^2 e^{-Nu(\alpha)} } { (1-\pi) (\alpha N+1) }
\label{equation:PEsubMDSbounded}
\end{equation}
where $u(\alpha)$ is defined as
\begin{equation}
u(\alpha)=\left\{ \begin{array}{ll}
           0 & \mbox{for $\alpha \leq \pi$}\\
             & \\
           \alpha \log \left( \dfrac{\alpha(1-\pi)}{\pi (1-\alpha)} \right) & \\
           - \log \left( \dfrac{1-\pi}{1-\alpha} \right) & \mbox{for $\pi < \alpha \leq 1$.} \end{array} \right.
\label{equation:ChernoffBoundDiscrete}
\end{equation}
with the $\log$ functions computed in the Neperian base.

Using equation~\eqref{equation:RvsAlpha}, the MDS coding error
exponent, $u(.)$, can be expressed in terms of $R$ instead of
$\alpha$. In~\eqref{equation:RvsAlpha}, $K$ should be an integer,
and we should have $q+1 \geq N$ for a feasible MDS code. Thus, the
finest resolution of rates achievable by a single MDS codebook would
be $R=\frac{i}{q+1} \log q$ for $i=1,2,\dots,q$. Of course, it is
also possible to achieve the rates in the intervals $\frac{i}{q+1}
\log q < R < \frac{i+1}{q+1} \log q$ by time sharing between two MDS
codebooks of sizes $[q+1,i]$ and $[q+1,i+1]$. However, in such
cases, the smaller error exponent belonging to the codebook of the
size $[q+1,i+1]$ dominates. Therefore, $u(R)$ will have a stepwise
shape of the form
\begin{equation}
u(R)=\left\{ \begin{array}{ll}

             0 & \hspace{-0.7cm} \mbox{for } 1-\pi \leq \tilde{r} \\
               & \hspace{-0.7cm} \\
             - \tilde{r} \log \dfrac { \left( 1-\pi \right) \left( 1-\tilde{r} \right) } {\tilde{r} \pi }  & \\
             - \log \dfrac{\pi} { 1-\tilde{r} }
               & \hspace{-0.7cm} \mbox{for } 0 < \tilde{r} \leq 1-\pi

             \end{array} \right.
\label{equation:uR}
\end{equation}
where $\tilde{r}$ is defined as
\begin{equation}
\tilde{r}= \dfrac{1}{q+1} \left\lceil \dfrac{(q+1)R}{\log q} \right\rceil
\end{equation}

\subsection{Random Coding Error Exponent of a Memoryless Erasure Channel}
\label{subsection:RandomCodeErrExpmemorylessErasureCh} 
It is interesting to compare the error exponent in \eqref{equation:uR} with the random
coding error exponent as described in~\cite{Gallager1968ErrExp}. This
exponent, $E_r(R)$, can be written as
\begin{equation}
E_r(R)=\displaystyle \max_{0 \leq \rho \leq 1} \left\{ -\rho R + \displaystyle \max_{\mathbf{Q}} E_0(\rho,\mathbf{Q}) \right\}
\label{equation:ErRMaximization}
\end{equation}
where $\mathbf{Q}$ is the input distribution, and
$E_0(\rho,\mathbf{Q})$ equals
\begin{equation}
\hspace{-0.25cm} E_0(\rho,\mathbf{Q}) = -\log \left( \displaystyle\sum_{j=0}^{q}
\left[ \displaystyle\sum_{k=0}^{q-1} Q(k)P(j|k)^{\dfrac{1}{1+\rho}} \right]^{1+\rho} \right).
\end{equation}
Due to the symmetry of the channel transition probabilities, the
uniform distribution maximizes \eqref{equation:ErRMaximization} over
all possible input distributions. Therefore, $E_0(\rho,\mathbf{Q})$
can be simplified as
\begin{equation}
E_0(\rho,\mathbf{Q}) = -\log \left( \dfrac{1-\pi}{q^\rho} + \pi \right).
\end{equation}
Solving the maximization~\eqref{equation:ErRMaximization}, gives us $E_r(R)$ as
\begin{equation}
E_r(R)= \left\{ \begin{array}{ll}
           -\log \dfrac{1-\pi+\pi q}{q} - r \log q  & \\
                                                        & \hspace{-3.2cm} \mbox{for } 0 \leq r \leq \dfrac{R_c}{\log q} \\
               & \\
           - r \log \dfrac{ \left( 1-\pi \right) \left( 1-r \right) } {r \pi  } - \log  \dfrac{\pi} { 1-r } & \\
                                                        & \hspace{-3.2cm} \mbox{for } \dfrac{R_c}{\log q} \leq r \leq 1-\pi
                \end{array} \right.
\label{equation:ErR}
\end{equation}
where $r=\frac{R}{\log q}$ , and $R_c=\frac{1-\pi}{1-\pi+\pi
q} \log q$ are the normalized and the critical rates, respectively.

Comparing \eqref{equation:uR} and~\eqref{equation:ErR}, we observe
that the MDS codes and the random codes perform exponentially the
same for rates between the critical rate and the capacity. However,
for the region below the critical rate, where the error exponent of
the random code decays linearly with $R$, MDS codes achieve a larger
error exponent. It is worth noting that this interval is negligible
for large alphabet sizes. Moreover, the stepwise graph of $u(R)$
meets its envelope as the steps are very small for large values of
$q$.

\begin{figure}
     \centering
     \subfigure[]{
           \label{fig:ErrorExp1}
           \includegraphics[width=0.45\textwidth]{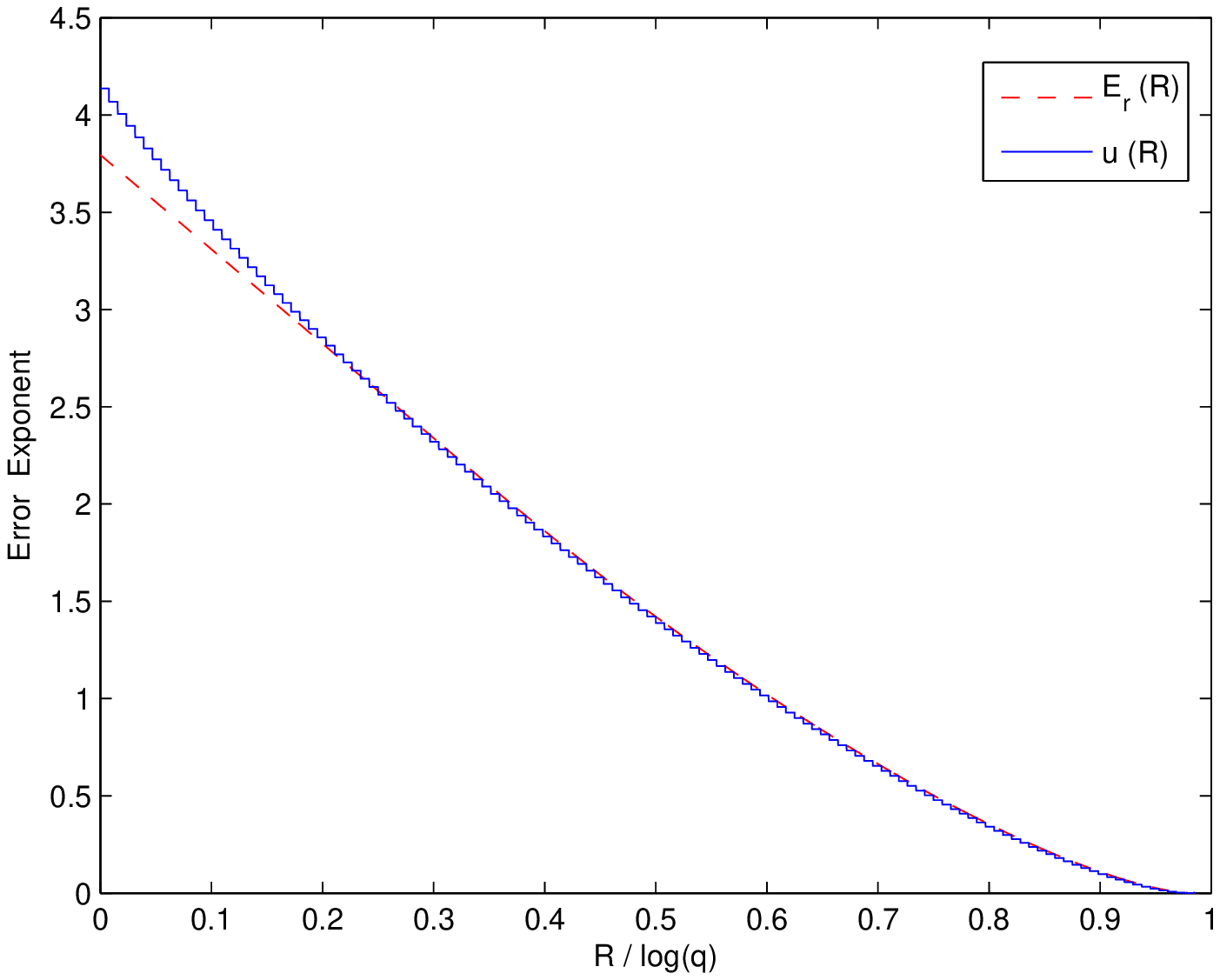}}
     \vspace{0.0in}
     \subfigure[]{
          \label{fig:ErrorExp2}
          \includegraphics[width=0.45\textwidth]{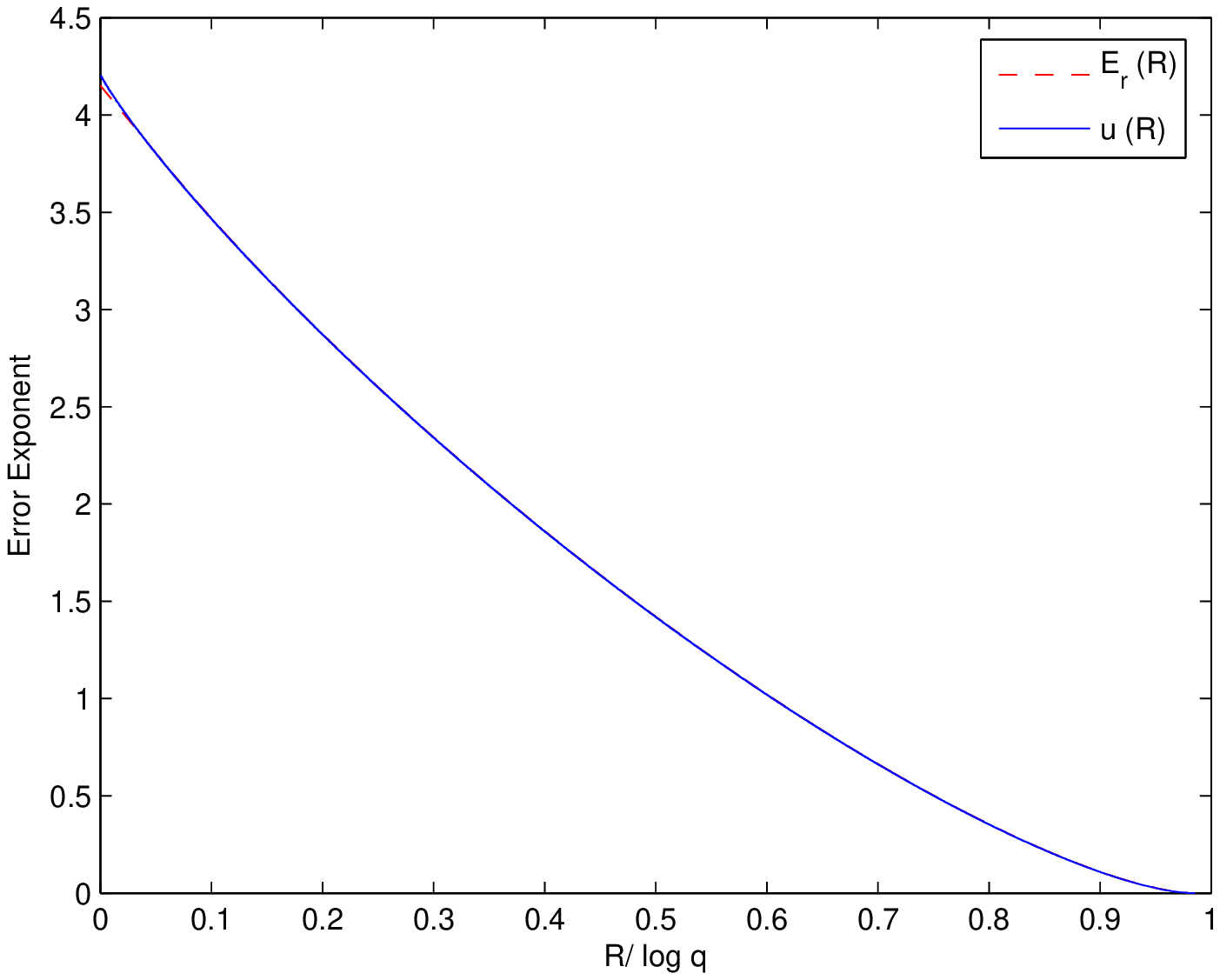}}
     \caption{Error exponents of random coding ($E_r(R)$) and MDS coding ($u(R)$) for a memoryless erasure channel with $\pi=0.015$, and (a): $q=128$, (b): $q=1024$.}
     \label{fig:ErrorExp}
\end{figure}

Figure~\ref{fig:ErrorExp} depicts the error exponents of random
codes and MDS codes for the alphabet sizes of $q=128$ and $q=1024$
over an erasure channel with $\pi=0.015$. As observed in
Fig.~\ref{fig:ErrorExp1}, $u(R)$ can be approximated by its envelope
very closely even for a relatively small alphabet size ($q=128$).
For a larger alphabet size (Fig.~\ref{fig:ErrorExp2}), the graph of
$u(R)$ almost coincides its envelope which equals $E_r(R)$ for the
region above the critical rate. Moreover, as observed in
Fig.~\ref{fig:ErrorExp2}, the region where MDS codes outperform
random codes becomes very small even for moderate values of alphabet
size ($q=1024$).

\subsection{Linear Random Coding Error Exponent of a Memoryless Erasure Channel}
\label{subsection:LinearRandomCodeErrExpMemorylessErasureCh}
Maximum likelihood decoding of random codes generally has exponential complexity in terms of the block length ($N$). Linear random codes, on the other hand, have the advantage of 
polynomial decoding complexity (assuming maximum likelihood decoding) over any arbitrary erasure channel~\cite{Dumer1994}. In a linear codebook of size $[N,K]$, any 
codeword, $\mathbf{c}$, can be written as $\mathbf{c}=\mathbf{b}\mathbf{G}$, where $\mathbf{b}$ is a row vector of length $K$, and indicates the information symbols. $\mathbf{G}$ is 
the generator matrix of size $K\times N$. In the case of a linear random codebook, every element in $\mathbf{G}$ is generated independently according to a 
distribution $\mathbf{Q}$~\cite{Pierce1967,Barg2002}. For a memoryless erasure channel, due to the symmetry of the channel transition probabilities, the uniform distribution is applied 
to generate $\mathbf{G}$.

Here, we describe a suboptimal decoder with polynomial complexity for decoding of linear block codes over erasure channels. This decoder is a slightly modified version of the 
optimum (maximum likelihood) decoder in~\cite{Dumer1994}. In case that less than $K$ symbols are received correctly, a decoding error is declared. When $K$ or more correct symbols 
are received, the decoder determines the information vector $\mathbf{b}$ (and the transmitted codeword $\mathbf{c}$) by constructing a new matrix called the \textit{reduced generator 
matrix}, $\mathbf{\tilde{G}}$. $\mathbf{\tilde{G}}$ consists of the columns in $\mathbf{G}$ whose corresponding symbols are received correctly. Thus, if the erasure identifier 
vector $\mathbf{e}$ has the weight of $w(\mathbf{e})=m \leq N-K$, $\mathbf{\tilde{G}}$ would have the size of $K \times (N-m)$. Then, the decoder computes the row or column rank 
of $\mathbf{\tilde{G}}$. If this rank is less than $K$, a decoding error is reported. In case the rank is equal to $K$, the information symbol vector can be decoded uniquely by 
solving $\mathbf{b}\mathbf{\tilde{G}}=\mathbf{\tilde{y}}$. In this case, $\mathbf{\tilde{y}}$ is the \textit{reduced received vector} consisting of the correctly received symbols only.

%
%

Using the described suboptimal decoder, the probability of error is the probability that the rank of $\mathbf{\tilde{G}}$ is less than $K$. Thus, the 
probability of error conditioned on an erasure vector of weight $w(\mathbf{e})=m$ can be written as~\cite{Didier2006}
\begin{equation}
\mathbb{P} \left\{ \mbox{error} | w(\mathbf{e})=m \right\} = 1- \prod_{i=N-m-K+1}^{N-m} \left(1-\dfrac{1}{q^i} \right).
\end{equation}
We bound the above probability as 
\begin{eqnarray}
\mathbb{P} \left\{ \mbox{error} | w(\mathbf{e})=m \right\} & \leq & 1-\left(1-\dfrac{1}{q^{N-m-K+1}} \right)^{K} \nonumber \\
                                                           & \stackrel{(a)}{\leq} & \dfrac{K}{q^{N-m-K+1}}
\label{equation:PeConditionalBound}
\end{eqnarray}
where $(a)$ follows from Bernoulli's inequality~\cite{Mitrinović1970} and the assumption that $w(\mathbf{e})=m \leq N-K$. The total probability 
of error is written as 
\begin{eqnarray}
P_{E,s\:\!\!u\:\!\!b}^{l\!i\!n} & = & \sum_{i=0}^{K-1} P_i + \sum_{i=K}^{N} P_i~\mathbb{P} \left\{ \mbox{error} | w(\mathbf{e})=N-i \right\} \nonumber \\
                                & \stackrel{(a)}{\leq} & \sum_{i=0}^{K-1} P_i + \sum_{i=K}^{N} \dfrac{KP_i}{q^{i-K+1}} \nonumber \\
                                & = & \sum_{i=0}^{K-2} P_i + Q_{K-1} + K \sum_{i=K}^{N} Q_i
\label{equation:PeSubLinBound}
\end{eqnarray}
where $P_i$ denotes the probability that $i$ symbols are received correctly as defined in 
subsection~\ref{subsection:ErrExpMDSmemorylessErasureCh}, and $Q_i=\dfrac{P_i}{q^{i-K+1}}$. $(a)$ follows 
from~\eqref{equation:PeConditionalBound}.

We define $i_0$ as $i_0=\frac{(N+1)(1-\pi)}{1-\pi+q\pi}$. Of course, $i_0$ is not necessarily an integer. For the case where $i_0 \leq K$, similar to equation~\eqref{equation:PiPi_1}, we can write
\begin{eqnarray}
\dfrac{Q_i}{Q_{i-1}}=\dfrac{(N-i+1)(1-\pi)}{qi\pi} \leq 1 & \mbox{for } i=K,\cdots,N. 
\label{equation:QiQi_1}
\end{eqnarray}
Thus, $Q_i$'s are decreasing, and we have 
\begin{eqnarray}
P_{E,s\:\!\!u\:\!\!b}^{l\:\!i\:\!n} & \stackrel{(a)}{\leq} & \sum_{i=0}^{K-1} P_i + K (N-K+1) Q_{K-1} \nonumber \\
                                & \stackrel{(b)}{\leq} & (N-K+2) K P_{K-1}  \nonumber \\
                                & \stackrel{(c)}{\leq} & \dfrac{\pi K^2 (N-K+2)}{(1-\pi)(N-K+1)} e^{-NE_r(R)} \nonumber \\
                                & = & \dfrac{\pi N^2 r^2 (N-Nr+2)}{(1-\pi)(N-Nr+1)} e^{-NE_r(R)}
\label{equation:AboveRc}
\end{eqnarray}
where $(a)$ follows from~\eqref{equation:QiQi_1} and~\eqref{equation:PeSubLinBound}. $(b)$ results from~\eqref{equation:PiPi_1}, and $(c)$ is 
based on~\eqref{equation:BinomBound} and~\eqref{equation:RvsAlpha}. The condition $i_0 \leq K$ can also be rewritten 
as $\frac{R_c}{\log q}\left(1+\frac{1}{N}\right) \leq r$ where $r=\frac{R}{\log q}$ as in~\eqref{equation:ErR}.

For the case where $K < i_0$, according to equation~\eqref{equation:QiQi_1}, the series of 
$\left\{ Q_i \right\}_{i=K-1}^{N}$ has its maximum 
at $i^{\star}=\lfloor i_0 \rfloor \geq K$. Thus, we have
\begin{eqnarray}
P_{E,s\:\!\!u\:\!\!b}^{l\:\!i\:\!n} & \stackrel{(a)}{\leq} & \sum_{i=0}^{K-1} P_i + K (N-K+1) Q_{i^{\star}} \nonumber \\
                                    & \stackrel{(b)}{\leq} & (N-K+2) K Q_{i^{\star}}  \nonumber \\
                                    & \stackrel{(c)}{\leq} & (N-K+2) K \exp \left( -N \left\{ \dfrac{i^{\star}}{N} \log \dfrac{ \dfrac{i^{\star}}{N}q\pi }{ \left(1-\dfrac{i^{\star}}{N}\right)\left(1-\pi\right) } - \log \dfrac{\pi}{1-\dfrac{i^{\star}}{N}} - \dfrac{K}{N} \log q \right\}\right) \nonumber \\
                                    & \leq & (N-K+2) K \exp \left( -N \left\{ \dfrac{i_0-1}{N} \log \dfrac{ \dfrac{i_0-1}{N}q\pi }{ \left(1-\dfrac{i_0-1}{N}\right)\left(1-\pi\right) } - \log \dfrac{\pi}{1-\dfrac{i_0}{N}} - \dfrac{K}{N} \log q \right\}\right) \nonumber \\
                                    & \stackrel{(d)}{=} & (N-rN+2)Nr e^{-Nv(R,N)}
\label{equation:BelowRc}
\end{eqnarray}
where $\exp(x)=e^{x}$, and $v(R,N)$ is defined as below  
\begin{eqnarray}
v(R,N) & = & \dfrac{N(1-\pi)-\pi q}{N(1-\pi+\pi q)} \log \dfrac{N(1-\pi)-\pi q}{(N+1)(1-\pi)} - \log \dfrac{\pi N(1-\pi+\pi q)}{N\pi q-1 +\pi} -R \nonumber \\
       & = & -\log \dfrac{1-\pi+\pi q}{q} - R + O(\dfrac{1}{N})~q.
\label{equation:vRN}
\end{eqnarray}
In~\eqref{equation:BelowRc}, $(a)$ follows from~\eqref{equation:QiQi_1} and~\eqref{equation:PeSubLinBound}, and $(b)$ results from~\eqref{equation:PiPi_1}. $(c)$ is 
based on~\eqref{equation:BinomBound}, and can derived similar to~\eqref{equation:PEsubMDSbounded}. $(d)$ follows 
from~\eqref{equation:RvsAlpha}. Combining~\eqref{equation:AboveRc} and~\eqref{equation:BelowRc} results in
\begin{equation}
P_{E,s\:\!\!u\:\!\!b}^{l\:\!i\:\!n} \leq \left\{ \begin{array}{ll}
                                                  (N-rN+2)Nr \exp\left(-N \left\{ -\log \dfrac{1-\pi+\pi q}{q} - R + O(\dfrac{1}{N})~q \right\} \right) & \mbox{for } R < R_c\left(1+\frac{1}{N}\right) \\
                                                     & \\
                                                  \dfrac{\pi N^2 r^2 (N-Nr+2)}{(1-\pi)(N-Nr+1)} \exp\left(-NE_r(R)\right) & \mbox{for } R \geq R_c\left(1+\frac{1}{N}\right) \\
                                                 \end{array} \right.
\end{equation}


\subsection{Exponential Optimality of Random Coding and Linear Random Coding}
\label{AsymOptRandCode} Using the sphere packing bound, it is shown
that random coding is exponentially optimal for the rates above the
critical rate over channels with relatively small alphabet sizes ($q
\ll N$)~\cite{Gallager1968SPB,Gallager1973}. In other words, we know that
\begin{equation}
P_{E,M\!L}^{r\:\!\!a\:\!\!n\:\!\!d} \doteq e^{-NE_r(R)}
\end{equation}
where the notation $\doteq$ means $\displaystyle\lim_{N->\infty}-\frac{\log P_{E,M\!L}^{r\:\!\!a\:\!\!n\:\!\!d}}{N}=E_r(R)$. However, the sphere packing 
bound is not tight for the channels whose alphabet size, $q$, is comparable
to the block length. Here, based on Proposition~I and the results of
section~\ref{section:ErrExpMDSrandom}, we prove the exponential
optimality of random coding and linear random coding over the erasure channels for 
all block sizes (both $N \geq q+1$ and $N<q+1$).

The average decoding error probability for an ensemble of random codebooks with the
maximum-likelihood decoding can be upper bounded as
\begin{equation}
P_{E,M\!L}^{r\:\!\!a\:\!\!n\:\!\!d} \stackrel{(a)}{\leq} e^{-NE_r(R)} \stackrel{(b)}{=} e^{-Nu(R)}
\label{equation:PEMLrandupperb}
\end{equation}
where $(a)$ follows from~\cite{Gallager1968ErrExp}, and $(b)$ is valid
only for rates above the critical rate according
to~\eqref{equation:uR} and~\eqref{equation:ErR}. The similar upper-bound for $P_{E,s\:\!\!u\:\!\!b}^{l\:\!i\:\!n}$ is given in~\eqref{equation:AboveRc}.

We can also lower bound $P_{E,M\!L}^{r\:\!\!a\:\!\!n\:\!\!d}$ and $P_{E,s\:\!\!u\:\!\!b}^{l\:\!i\:\!n}$ as
\begin{eqnarray}
P_{E,M\!L}^{r\:\!\!a\:\!\!n\:\!\!d} & \hspace{-0.25cm} \stackrel{(a)}{\geq} & \hspace{-0.25cm} P_{E,M\!L}^{M\!DS} \nonumber \\
                                    & \hspace{-0.25cm} \stackrel{(b)}{\geq} & \hspace{-0.25cm} \left(1-\frac{1}{q}\right)  P_{E,s\:\!\!u\:\!\!b}^{M\!DS} \nonumber \\
                                    & \hspace{-0.25cm} \stackrel{(c)}{\geq} & \hspace{-0.25cm}
                                    \frac { \left(1-\frac{1}{q}\right) \pi r N e^{-Nu(R)} } { \left(1-\pi\right) \left(N+1\right) \left(\left(1-r\right) N+1\right) }
\label{equation:PEMLrandlowerb}
\end{eqnarray}
where $(a)$ follows from Proposition I and~\eqref{equation:PEMLlowerbound}, $(b)$ from
inequality~\eqref{equation:PEPEboundedML}, and $(c)$ from
inequality~\eqref{equation:PEsubMDSbounded}. The inequality in~\eqref{equation:PEMLrandlowerb} remains 
valid if $P_{E,M\!L}^{r\:\!\!a\:\!\!n\:\!\!d}$ is replaced by $P_{E,s\:\!\!u\:\!\!b}^{l\:\!i\:\!n}$.

Combining~\eqref{equation:PEMLrandupperb}
and~\eqref{equation:PEMLrandlowerb} guarantees that both the
upper-bound and the lower-bound on
$P_{E,M\!L}^{r\:\!\!a\:\!\!n\:\!\!d}$ are exponentially tight, and
the decaying exponent of $P_{E,M\!L}^{r\:\!\!a\:\!\!n\:\!\!d}$
versus $N$ is indeed $u(R)$. Combining~\eqref{equation:AboveRc} and~\eqref{equation:PEMLrandlowerb} proves 
the same result about the exponent of $P_{E,s\:\!\!u\:\!\!b}^{l\:\!i\:\!n}$ versus $N$. Moreover, we can write
\begin{eqnarray}
P_{E,M\!L}^{M\!DS} \stackrel{(a)}{\leq} P_{E,M\!L}^{r\:\!\!a\:\!\!n\:\!\!d} \stackrel{(b)}{\leq} \frac { (1-\pi) (N+1) (N-rN+1) } { \left(1-\frac{1}{q}\right) \pi r N } P_{E,M\!L}^{M\!DS} \nonumber \\
P_{E,M\!L}^{M\!DS} \stackrel{(a)}{\leq} P_{E,s\:\!\!u\:\!\!b}^{l\:\!i\:\!n} \stackrel{(c)}{\leq} \frac { \pi N r (N+1) (N-rN+2) } { \left(1-\frac{1}{q}\right) } P_{E,M\!L}^{M\!DS}
\label{equation:PEMLrandboundedopt}
\end{eqnarray}
where $(a)$ follows from Proposition I and~\eqref{equation:PEMLlowerbound}, and $(b)$ results from
inequalities~\eqref{equation:PEMLrandupperb} and
\eqref{equation:PEMLrandlowerb}. $(c)$ is 
based on~\eqref{equation:PEPEboundedML},~\eqref{equation:PEsubMDSbounded}, and~\eqref{equation:BelowRc}. Since the coefficients of
$P_{E,M\!L}^{M\!DS}$ in~\eqref{equation:PEMLrandboundedopt} do not include any exponential
terms, it can be concluded that for rates above the critical rate,
both random codes and linear random codes perform exponentially the same as MDS codes, which are
already shown to be optimum.

\section{Conclusion}
\label{section:Conclusion} 
Performance of random codes, linear random codes, and MDS codes
over an erasure channel with a fixed, but large alphabet size is
analyzed. We proved that MDS codes minimize the probability of
decoding error (using maximum-likelihood decoding) over any erasure
channel (with or without memory). Then, the decoding error
probability of MDS codes, random codes, and linear random codes are 
bounded by exponential terms, and the corresponding exponents are 
compared. It is observed that the error exponents are identical over a 
wide range of rates. Knowing MDS codes are optimum, it is concluded that 
both random coding and linear random coding are exponentially optimal over 
a memoryless erasure channel for all block sizes (whether $N \geq q+1$ or $N<q+1$).
\section*{Acknowledgments}
\label{section:Acks}
The authors would like to thank Dr. Muriel Medard and Dr. Amin Shokrollahi 
for their helpful comments and fruitful suggestions to improve this work. 
\bibliographystyle{IEEEtran}
\bibliography{isit2008_v5}

\end{document}